\documentclass[12pt]{article}

\usepackage{epsfig}
\usepackage{amssymb}
\usepackage{amsmath}

\newcommand{\bea}{\begin{eqnarray}}
\newcommand{\eea}{\end{eqnarray}}

\begin{document}

\title{\bf Scattering amplitudes at strong coupling for \\ $4K$ gluons}

\author{Gang Yang\thanks{E-mail: g.yang@qmul.ac.uk} }
\date{}

\maketitle

\centerline{\it Centre for Research in String Theory}
\centerline{\it Department of Physics, Queen Mary, University of
London}
\centerline{\it Mile End Road, London, E1 4NS, United Kingdom}

\begin{abstract}

In this paper we study the scattering amplitudes at strong coupling
for the case where the number of gluons is a multiple of four. This
is an important missing piece in \cite{Alday:2010vh}. The tricky
point for $n=4K$ is that there is some accidental degeneracy in such
case. We explain this point in detail and show that a non-trivial
monodromy around infinity was developed by the world-sheet
coordinate transformation appearing in the computation. It turns out
that besides solving the $Y$ system, we also need to calculate $T$
functions to compute the full amplitudes. We show that the $T$
functions can be derived by taking a limit of $Y$ functions of a
higher-point case. As a check, we obtain the known result of
eight-point in $AdS_3$ in \cite{Alday:2009yn}.

\end{abstract}

\section{Introduction}

Scattering amplitudes are central quantities in quantum field
theory. The knowledge of their behavior at higher loops and at
strong coupling may be instrumental in understanding the problems
such as quark confinement or quantum gravity. While it is very hard
to do such calculations in QCD or in gravitational theories, many
significant developments in past several years have shown that it
may be possible to have a non-perturbative understanding of
$S$-matrix in ${\cal N}=4$ SYM.

\vskip 0.08in

Based on the explicit perturbative calculation, Bern, Dixon and
Smirnov proposed a non-perturbative conjecture for planar MHV
amplitudes in ${\cal N}=4$ SYM, for all number of gluons up to all
loops \cite{BDS}. This is now well-known as BDS ansatz. The idea was
also indicated before in \cite{ABDK}. This ansatz was supported by
the later calculation of two-loop five-point amplitude and
four-point amplitude up to five loops \cite{4p4l,4p4l2}. The
(generalized) unitarity method plays an essential role for doing the
higher-loop calculation \cite{BDDK,Britto:2004nc}.

At strong coupling, by using AdS/CFT duality \cite{Maldacena}, a
recipe for calculating scattering amplitudes was also proposed by
Alday and Maldacena \cite{AM}. The problem is reduced to calculating
the area of minimal surfaces in $AdS_5$ ending on a null polygon at
the boundary, where the shape of the polygon is determined by the
momenta of external gluons. Due to the similar prescription for
Wilson loop \cite{Maldacena:1998im,Rey:1998ik}, this indicated that
there may be a duality between amplitudes and Wilson loops at weak
coupling, which was soon proved to be true at one loop for general
$n$ points, and for four and five points at two loops
\cite{Drummond123, Brandhuber:2007yx, Drummond3}.

At the same time, the BDS ansatz was also questioned by the study of
amplitudes at strong coupling for large number of gluons \cite{AM2}.
Later the explicit weak coupling two-loop six-point calculations
showed that the BDS ansatz is incorrect while the duality between
amplitudes and Wilson loops is still true \cite{two-loop-six-gluon,
two-loop-six-Wilson-loop}. On the other hand, the BDS ansatz gives
the correct conformal anomaly of Wilson loops \cite{Drummond3}%
\footnote{At one loop level, the conformal anomaly has been proved
for general $n$-point amplitudes \cite{Brandhuber:2009xz}.}.
Therefore, under the assumption of the amplitude/Wilson loop
duality, the difference between BDS ansatz and the true result
should be a (dual) conformal invariant quantity, which is usually
referred to as the ``remainder function". To fully understand planar
MHV amplitudes, the main problem is to understand
this mysterious remainder function%
\footnote{We should mention that there are other very important
problems about understanding non-MHV (and also non-planar)
amplitudes. The dual conformal supersymmetry \cite{Drummond:2008vq},
fermionic T-duality \cite{Berkovits:2008ic,Beisert:2008iq}, and
Grassmannian integral \cite{ArkaniHamed:2009dn,Mason:2009qx} are
some important developments along these lines. }. A numerical
program for calculating two-loop Wilson loop was developed in
\cite{Anastasiou:2009kna}, and some properties of the remainder
functions beyond six-point were studied in \cite{Brandhuber:2009da}.
The analytic calculation of remainder function for six-point was
also performed in \cite{remainder-six, remainder-six-simple}.

\vskip 0.08in

Unlike at weak coupling, the calculation of amplitudes at strong
coupling is a geometric minimal surface problem. For the simplest
four-point case \cite{AM}, the solution of the minimal surface was
obtained by some guess, or by doing conformal transformations to a
cusp solution \cite{Kruczenski:2002fb}. But it is very hard to find
solutions for higher-point cases. Remarkably, in a series of papers
\cite{Alday:2009yn,Alday:2009dv,Alday:2010vh}, Alday, Maldacena and
collaborators developed a method which makes it possible to
calculating the area of minimal surface with general null polygonal
boundary conditions, where the integrability of the system plays an
essential role \cite{Mandal:2002fs,Minahan:2002ve,Bena:2003wd}.
Using this method, one can calculate the area directly without the
need of constructing the explicit solution of the minimal surface.
Let us briefly mention some key steps here.

The first important trick is the Pohlmeyer reduction
\cite{Pohlmeyer, Barbashov:1982qz, DeVega:1992xc} (see also
\cite{Jevicki:2007aa,{Grigoriev:2008jq},Miramontes:2008wt,Dorn:2009gq}
for some recent developments). By using this reduction, solving the
classical string equations and the Virasoro constraints becomes
solving a Hitchin system (with a $Z_4$ projection). A very important
fact for Hitchin system is that the equations can be promoted by
introducing a spectral parameter $\zeta$. This turns out to be
instrumental for solving the problem. In particular, by introducing
this auxiliary parameter, the cross ratios can be promoted to a
function of spectral parameter. The functional relations between
cross ratios can be organized in a framework of the so called $Y$
system \cite{Yang:1968rm,Zamolodchikov:1989cf} (see also
\cite{Dorey:1998pt, Dorey:1999uk}), where $Y$ functions are the
cross ratios. Under this framework, one can write a set of integral
equations, where the boundary conditions can be very nicely embedded
via WKB approximation at large and small $\zeta$
\cite{Gaiotto:2008cd,Gaiotto:2009hg}, and finally, the non-trivial
part of the area can be expressed as the free energy of the $Y$
system.

\vskip 0.08in

While the above prescription works well for the case where the
number of gluons is odd, it can not be applied directly to the case
where the number of gluons $n$ is even%
\footnote{This is because the intersection form of the Riemann
surface appearing in the calculation is only invertible for the case
where $n$ is odd \cite{Alday:2010vh}.}. For such cases, one may
obtain the result by taking a limit of $(n\!+\!1)$-point case. This
is relatively trivial when $n$=$4K$+2 \cite{Alday:2010vh}. However,
the calculation is much more subtle when $n$=$4K$, i.e. the number
of gluons is a multiple of four. As we will show, such cases are
special in that a world-sheet coordinate transformation appearing in
the computation develops a non-trivial monodromy around infinity.
This makes the calculation of the so called cutoff part and periods
part much more non-trivial. In the simple $AdS_3$ case, a
prescription was given in \cite{Alday:2009yn}, but a full
prescription for $AdS_5$ case is still unknown. This is the problem
that we consider in this paper.

We will provide a general prescription for the computation of the
cutoff part. In $n\!\neq\! 4K$ cases, the cutoff part is trivial and
can be uniquely written in terms of only adjacent kinematic
invariants. But in the cases of $n$=$4K$, these are the main
complications. We will show the problem can be solved by introducing
two extra equations which involve non-adjacent kinematic invariants.
These equations also involve the so called $T$-functions. The parts
that depend on the $T$-functions are defined as extra part, while
the remaining parts are defined as BDS-like part. We show that the
$T$ functions can be calculated as a limit of $Y$ functions. Our
prescription reproduces the known $AdS_3$ result, which provides a
strong check for the consistency and validity of the method.

\bigskip

The paper is organized as follows. In section
\ref{section-amplitudes} we review the general structure of
amplitudes at strong coupling. In section \ref{section-cutoff} we
study in detail the origin of the subtly in $n$=$4K$ case and
calculate the cutoff part for such case. In section \ref{section-T},
we calculate $T$ functions as a limit of $Y$ functions. In section
\ref{section-periods} we make a conjecture for the periods part. We
present the explicit eight-point result in section \ref{section-8p}.
Section \ref{section-dis} contains some discussions. We give a brief
summary for the $Y$ system in the appendix.

\section{Structure of amplitudes at strong coupling
\label{section-amplitudes}}

The general structure of amplitudes at strong coupling can be given
as
\bea A  & = &  A_{\rm div} + A_{\rm BDS-like} + A_{\rm extra} +
A_{\rm periods} + A_{\rm free} ~. \eea
The free and periods parts are basically the parts that can be
calculated via $Y$ system \cite{Alday:2010vh}. The cutoff part is
constituted of $A_{\rm div}$, $A_{\rm BDS-like}$ and $A_{\rm extra}$
parts. As we will see in next section, the extra part appears only
in the $n$=$4K$ case.  We emphasize that although the free and
periods parts may be the most non-trivial part of the amplitudes,
the cutoff part also contains important physical information. For
example, for four and five-point cases in particular, the cutoff
part gives the whole result, therefore contains the whole physics.

Let us look at the origin of each part more closely.%

By Pohlmeyer reduction, the area of the minimal surface can be
written as
\bea A = 2 \int d^2 z {\rm Tr}[\Phi_z \Phi_{\bar z}] ~, \eea
where $\Phi$ is a component of flat connection of the corresponding
Hitchin system. The boundary conditions of the problem require that
${\rm Tr}[\Phi_z \Phi_{\bar z}] \rightarrow (P(z)\bar P(\bar
z))^{1/4}$ for large $z$, therefore one can regularize the area by
subtracting the asymptotic divergent part as
\bea A_{\rm free}
= 2 \int d^2 z {\rm Tr}[\Phi_z \Phi_{\bar z}] - 2 \int d^2 w ~,
\qquad d w = P(z)^{1/4} d z ~, \eea
where for $n$-point $P(z)$ is a polynomial of degree $n$-4. This
part is called free part since it turns out to be the free
energy of the corresponding $Y$ system \cite{Alday:2010vh}%
\footnote{Notice that in \cite{Alday:2010vh} the method used to
deriving $A_{\rm free}$ is only valid when $n$ is odd, since the
intersection form of the Riemann surface is singular in other cases.
One can argue that by starting from odd $n$ result and taking a
large $m$ limit, we have the same expression of free part for all
other cases. }
\bea A_{\rm free} = \sum_s { m_s \over 2\pi}
\int_{-\infty}^{+\infty}d\theta \cosh\theta \log \left[(1+
Y_{1,s})(1+ Y_{3,s})(1+ Y_{2,s})^{\sqrt{2}}\right] ~. \eea
The area can then be written as
\bea A = A_{\rm free} + 2 \int d^2 w ~.  \eea

The second term is still divergent, therefore needs regularization.
It is convenient to introduce a cutoff surface $\Sigma$ and consider
the integral $\int_\Sigma d^2 w$
\footnote{In the weak coupling calculation, dimensional
regularization is more convenient. At strong coupling as a
geometrical problem the cutoff regularization appears to be very
natural. This cutoff regularization may be related to the off-shell
and Higgs regularization at weak coupling \cite{Drummond:2006rz,
Dorn:2008dz, Alday:2009zm}. }.
Notice the surface $\Sigma$ still contains complicated branch cut
information which is given by the polynomial $P(z)$. To simplify the
problem, we can introduce another surface $\Sigma_0$ with the same
cutoff but with simpler internal structure. Then we can separate the
second term further into two parts
\bea 2 \int_\Sigma d^2 w = A_{\rm periods} + A_{\rm cutoff} ~, \eea
where
\bea A_{\rm periods} = 2 \int_\Sigma d^2 w - 2 \int_{\Sigma_0} d^2 w
~, \qquad A_{\rm cutoff} = 2 \int_{\Sigma_0} d^2 w ~. \eea
While $n\neq 4K$ we can define $\Sigma_0$ corresponding to a
polynomial whose zeros are all degenerate at the origin. Then the
periods part can be defined explicitly as
\bea A_{\rm periods} = 2 \int d^2 z \left( [ P(z) \bar P(\bar z)]
^{1/4} - |z|^{n/2-2} \right) ~, \eea
which can be expressed in terms of periods around cycles of the
Riemann surface%
\footnote{The Riemann surface appearing here is defined as algebraic
curves which is related to the polynomial $P(z)$. For $AdS_5$ case,
the Riemann surface is defined as $x^4$=$P(z)$ which is a quadruple
branch cover of Riemann sphere. While for $AdS_3$ case, it is only a
double branch cover defined by $x^2$=$p(z)$ (where $P(z) = p(z)^2$
in such case).}, therefore explains why it's called periods part.
Using $Y$ system, this part can be calculated together with the free
part when $n$ is odd. For the case of $n$=$4K$+2, the result can
also be obtained by taking a large mass limit of $n$=$4K$+3 case
\cite{Alday:2010vh}. However the $n=4K$ case is more tricky, we will
discuss this part in section \ref{section-periods}.

\vskip 0.1in

The remaining part is the cutoff part. If the number of gluons is
not a multiple of four, the calculation is very simple. There is no
extra part in such cases. And besides the universal divergent part,
the BDS-like part turns out to be the unique solution of the dual
conformal Ward identity which is expressed in terms of only adjacent
kinematic invariants $x_{i,i+2}^2$. Explicitly, for $n=4K+2$, we
have
\bea A_{\rm BDS-like} = -{1\over 8} \sum_{i=1}^n \Big( \ell_i^2 +
\sum_{k=0}^{2K} \ell_i \ell_{i+1+2k} (-1)^{k+1} \Big) ~,
\label{bdslike4k2} \eea
while $n=4K+1,~ 4K+3$ we have
\bea A_{\rm BDS-like} = -{1\over 4} \sum_{i=1}^n \Big( \ell_i^2 +
\sum_{k=0}^{2K} \ell_i \ell_{i+1+2k} (-1)^{k+1} \Big) ~,
\label{bdslike4k13} \eea
where
\bea \ell_i \equiv \log x_{i,i+2}^2 ~. \eea
However, when $n\!=\!4K$, the calculation of cutoff part becomes
much more complicated, due to the existence of a monodromy around
infinity which we will discuss in detail in next section.

\section{Cutoff part \label{section-cutoff}}

We calculate the cutoff part in this section. We first review the
embedding coordinate and cutoff regulator. Then we discuss why the
calculation is tricky for the $n$=$4K$ case from various points of
view. We show that one can calculate the cutoff part for such case
by introducing two new equations involving non-adjacent kinematic
invariants, and also $T$ functions which give the extra part.

\subsection{Embedding coordinates and cutoff regulator}

It is convenient to work in the embedding coordinates of $AdS_5$
space
\bea X\cdot X \equiv - X^+ X^- + X^\mu X_\mu = -1~, \qquad \mu =
0,1,2,3, \eea
where
\bea X^\pm = X^{-1}\pm X^4~. \eea
The boundary of $AdS_5$ space can be defined as $ \hat X= X /R$, by
taking $R \rightarrow \infty$
\bea {\hat X}^2 = - {\hat X}^+ {\hat X}^- + {\hat X}^\mu {\hat
X}_\mu = 0 , \quad {\hat X}\sim \lambda {\hat X} ~. \eea
The relation between embedding coordinates and Poincar$\acute{\rm
e}$ coordinates is
\bea x_\mu = {X_\mu \over X_+}, \qquad  r = {1\over X^+} = {1\over
\hat X^+ R} ~, \eea
where in Poincar$\acute{\rm e}$ coordinates the boundary is defined
at $r\rightarrow 0$, which is consistent with taking
$R\rightarrow\infty$.

\vskip 0.1in

To impose the cutoff, we need to understand the asymptotic behavior
of the minimal surface. An important trick to impose the boundary
condition is to change the world-sheet coordinate from original $z$
coordinate to $w$ coordinate, via $d w = P(z)^{1/4} d z$
\cite{Alday:2009dv}. In the new $w$ coordinate, every cover
of $w$-plane contains only four cusps%
\footnote{Since every cover of $w$-plane contains four cusp, the
asymptotic behavior should be the same as the four-cusp solution
which is well understood. This is why it's easy to impose boundary
condition by using $w$ coordinate.}, and the minimal surface with
$n$ cusps covers the $w$-plane $n/4$ times. Due to the non-trivial
polygonal boundary condition, the solution of the minimal surface
has different asymptotic behaviors near different cusps, which can
be described by the so call ``Stokes phenomenon"
\cite{Gaiotto:2008cd, Gaiotto:2009hg}. Each cusp corresponds to one
Stokes sector, and each stokes sector has one smallest solutions
$s_i$ that decay fastest to the boundary. Therefore, for every cover
of $w$ plane, we have four Stokes sectors and four smallest
solutions.

\vskip 0.1in

\begin{figure}[t]
\centerline{\includegraphics[height=6cm]{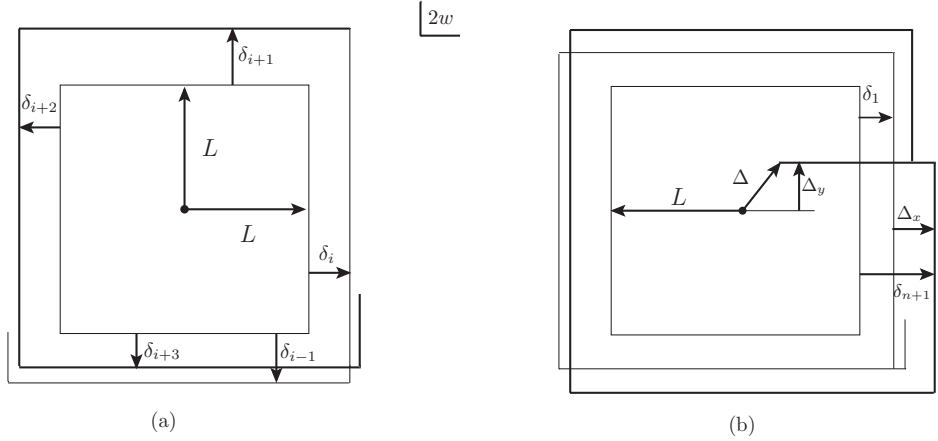} }
\caption{\it The cutoff of the surface $\Sigma_0$. Fig (a) shows a
portion of the surface in the $w$-plane. $L=-\log\epsilon_c$ is the
cutoff. $\delta_i = -\log {\hat X}_i^+$. The origin should be chosen
to be one of zeros of the polynomial $P(z)$. Fig (b) shows that for
$n=4K$ cases the surface is not closed. There is a formal monodromy
$\Delta = \Delta_x + i \Delta_y$, thus $\delta_{n+1} = \delta_1 +
\Delta_x$, $\delta_{n+2} = \delta_2 + \Delta_y$. The total area is
the sum of the area of various rectangles. Notice that we choose to
treat the first cusp in a special way. Half of it from $\delta_1$ at
the beginning, and half from the end of surface with $\delta_{n+1}$
which includes the effect of monodromy.}\label{fig-cutoff}
\end{figure}

Now we can regularize the surface. As in the usual way, we introduce
a cutoff for the radius of $AdS_5$
\bea r > \mu ~, \qquad {\rm or~equivalently} \quad   X^+ < {1 \over
\mu} ~. \eea
The asymptotic behavior of the solution near each cusp can be given in $w$-plane as%
\footnote{Notice that we have rotated the original $w$-plane by
$\pi/4$ to simplify the exponential function. They are actually
similar to the $u,v$-plane in \cite{Alday:2009yn}.}
\bea X_i^A \simeq {\hat X}_i^A \times \left\{ e^{2{\rm Re}[w]}, ~~
e^{2{\rm Im}[w]}, ~~ e^{-2{\rm Re}[w]}, ~~ e^{-2{\rm Re}[w]}
\right\}~.  \label{Xasymptotic}\eea
Therefore, the cutoff for the radius effectively becomes a cutoff
for the $w$-plane. For example for four consecutive cusps in one
cover of $w$-plane
\bea \begin{matrix} {\hat X}_i^+ e^{2{\rm Re}[w]} < {1 \over \mu} ~,
&& {\hat X}_{i+1}^+ e^{2{\rm Im}[w]} < {1 \over \mu} ~, \\ \\ {\hat
X}_{i+2}^+ e^{-2{\rm Re}[w]} > {1 \over \mu} ~, && {\hat X}_{i+3}^+
e^{-2{\rm Im}[w]} > {1 \over \mu} ~,
\end{matrix}  \eea
or equivalently
\bea \begin{matrix} 2 {\rm Re}[w] < L + \delta_i ~,  && 2 {\rm
Im}[w] < L + \delta_{i+1} ~, \\ \\ 2 {\rm Re}[w] > -(L +
\delta_{i+2}) ~, && 2 {\rm Im}[w] > -(L + \delta_{i+3}) ~,
\end{matrix}  \eea
where we have defined
\bea \delta_i \equiv - \log {\hat X}_i^+ ~, \qquad  L \equiv - \log
\mu \gg \delta_i ~. \eea
A portion of the regularized surface is shown in Figure
\ref{fig-cutoff}(a).

\vskip 0.1in

Besides using the $w$ coordinate for world-sheet, it is also
instrumental to use the spinor representation of $SO(2,4)$ for
target space. This was implied firstly from the study in the $AdS_3$
case \cite{Alday:2009yn}, where the technique is similar to the
spinor helicity formalism (see for example \cite{Dixon:1996wi,
Witten:2003nn}). In $AdS_5$ case, the spinor representation of
$SO(2,4)$ becomes the fundamental of $SU(2,2)$. Very interestingly,
this representation is equivalent to that of momentum twistor
variables which was first introduced at weak coupling by Hodges in
\cite{Hodges:2009hk} (see also \cite{Mason:2009qx}). The smallest
solutions $s_i$ of each Stokes sector play exactly the role of
momentum twistor variables. And we have the important relations
\bea x_{ij}^2 = { {\hat X}_i\cdot {\hat X}_j \over {\hat X}_i^+
{\hat X}_j^+ } ~, \qquad {\hat X}_i\cdot {\hat X}_j = \langle s_i
s_{i+1} s_j s_{j+1} \rangle ~, \qquad {\hat X}_i^{\alpha\beta} \sim
s_i^\alpha \wedge s_{i+1}^\beta ~. \label{xij2} \eea
These smallest solutions and their contractions are the basic block
of $Y$ system as review in Appendix \ref{appendix-TY}.

Notice we can rewrite (\ref{xij2}) as
\bea \delta_i + \delta_j = \ell_{ij} - \log ({\hat X}_i\cdot {\hat
X}_j) ~, \qquad \ell_{ij} \equiv \log x_{ij}^2 ~.
\label{non-adjacent} \eea
For adjacent case, they are simplified as
\bea \delta_i + \delta_{i+2} = \ell_i ~,  \label{eqell} \qquad
\ell_i \equiv \log x_{i,i+2}^2 ~, \eea
where we can use the normalization condition (\ref{s1234}), so that
$ {\hat X}_i\cdot {\hat X}_{i+2} = \langle s_i s_{i+1} s_{i+2}
s_{i+3} \rangle = 1$.

\subsection{Why $n=4K$ is special}

With the above preparation, we can now calculate the cutoff part. We
start from the $n\neq 4K$ case. In such case, the cutoff part is
simply given by summing over all rectangles of the surface as shown
in Figure \ref{fig-cutoff}(a). The whole contribution is
\bea A_{\rm cutoff} = {1\over 2} \sum_{i=1}^n (L + \delta_i)(L+
\delta_{i+1}) ~. \eea
This can be separated into a divergent part and a finite part as
\bea && A_{\rm cutoff} = A_{\rm div} + A_{\rm BDS-like} ~, \\
&& A_{\rm div} = {1\over 2} \sum_{i=1}^n \Big( L+{\delta_i +
\delta_{i+2} \over2} \Big)^2 ~,
\\ && A_{\rm BDS-like} = -{1\over 4} \sum_{i=1}^n \delta_i (\delta_i
+ \delta_{i+2} - 2\delta_{i+1} ) ~. \label{bdslike-delta}  \eea
Now we need to solve for $\delta_i$ in terms of the kinematic
variables. For the $n\neq4K$ case, it is enough to consider the
equations involving only adjacent kinematic invariant (\ref{eqell})
\bea \delta_i + \delta_{i+2} = \ell_i ~, \qquad i=1,2,\cdots, n ~.
\eea
We also impose the periodic condition $\delta_{i+n}=\delta_i$. Then
it is easy to solve these $n$ equations and express $\delta_i$ in
terms of $\ell_i$. By substituting the solution into
(\ref{bdslike-delta}), we obtain exactly the expression of cutoff
part (\ref{bdslike4k2}) and (\ref{bdslike4k13}).

\vskip 0.1in

However, the above prescription is no longer true when $n=4K$. In
particular the periodic condition is no longer allowed. This is
because the $n$ equations (\ref{eqell}) are decoupled into two sets:
one only involves odd indices, the other only involves even
indices%
\footnote{We emphasize that this degeneracy of equations only
appears when $n=4K$ and is in some sense the root that why it is
tricky for such cases.}
\bea \begin{matrix} \delta_{2k+1} + \delta_{2k+3} = \ell_{2k+1} &
\rightarrow & \delta_1 + \delta_3 = \ell_1 ~, & \cdots ~, &
\delta_{n-1} + \delta_{n+1} = \ell_{n-1} ~; \\ \\ \delta_{2k} +
\delta_{2k+2} = \ell_{2k} \qquad  & \rightarrow  & \delta_2 +
\delta_4 = \ell_2 ~, & \cdots ~, & \delta_{n} + \delta_{n+2} =
\ell_{n} ~. \qquad
\end{matrix} \label{eqdecouple} \eea
If we still impose the periodic condition $\delta_{n+i}=\delta_i$,
we would have
\bea && \ell_1 - \ell_3 + \ell_5 - \cdots + \ell_{n-3} - \ell_{n-1}
= 0 ~, \\ && \ell_2 - \ell_4 + \ell_6 - \cdots + \ell_{n-2} -
\ell_{n} = 0 ~, \eea
which is in general not true. Therefore, we have to break the
periodic condition and let
\bea \delta_{n+1} = \delta_1 + \Delta_x ~, \qquad \delta_{n+2} =
\delta_2 + \Delta_y ~, \label{deltaxy} \eea
by introducing two new variables $\Delta_{x,y}$.

In the $w$-plane, this non-periodic condition for $\delta_i$ means
that after going around the $w$-plane $n/4$ times, the origin of $w$
plane experiences a shift
\bea w \rightarrow w + \Delta ~, \qquad \Delta = \Delta_x + i
\Delta_y ~. \eea
This is illustrated in Figure \ref{fig-cutoff}(b).

\bigskip

This shift can also be understood from another point of view. Notice
that world-sheet coordinate transformation is defined as $w = \int
P(z)^{1/4} d z$. Since for $n$ points, the degree of polynomial
$P(z)$ is $n-4$, it is \textit{only} in the $n=4K$ case that there
is a single pole for
\bea P(z)^{1/4} \sim z^{n/4-1} + \cdots + {c \over z} + \cdots ~.
\eea
Therefore the cycle integral is non-zero around infinity in such
case. The means that the shift we impose above is actually the
monodromy around infinity in the $w$-plane
\bea \Delta \sim \oint_{\gamma^\infty} P(z)^{1/4} dz ~. \eea

\vskip 0.08in

By solving the equations (\ref{eqdecouple}) and (\ref{deltaxy}), one
can express the monodromy in terms of kinematic variable as
\bea \Delta_x = -\ell_1+\ell_3-\ell_5+\ell_7 - ... + \ell_{n-1} ~,
\qquad \Delta_y = -\ell_2+\ell_4-\ell_6+\ell_8 - ... + \ell_n ~.
\eea
It is interesting to see that this is equivalent to  the following
relation (by using (\ref{xij2}) and (\ref{s1234}))
\bea \Delta_x = \log \langle s_1 s_2 s_{n-1} s_n \rangle, \qquad
\Delta_y = \log \langle s_2 s_3 s_n s_{n+1} \rangle ~,
\label{deltaT} \eea
which is related to a $T$ function (\ref{Tads5})
\bea T_{2,n-4} = \langle s_0 s_1 s_{n-2} s_{n-1} \rangle^{[-n+2]} =
\big( e^{-\sqrt{2}({w_0\over\zeta}+\bar w_0 \zeta)} \big)^{[-n+2]}
~. \eea
The second equation was derived in \cite{Alday:2010vh}, where $w_0$
is called formal monodromy%
\footnote{There is also another parameter $\mu$ also contribute to
the formal monodromy in \cite{Alday:2010vh}, which is is related to
gauge connection, and has no relation with the discussion here.}.
%
%
\bea w_0 = (m_{n-5} + \sqrt{2} m_{n-6} + m_{n-7}) - (m_{n-9} +
\sqrt{2} m_{n-10} + m_{n-11}) + \cdots ~. \eea
Via monodromy, this provides one simple relation between mass
parameters and kinematic invariants.
%
%

\subsection{Cutoff part of $n\!=\!4K$ case}

The cutoff surface for the $n$=$4K$ case has the structure as shown
in Figure \ref{fig-cutoff}(b). To calculate the cutoff part, we sum
over all rectangles as in $n\neq 4K$ cases, but we also need to
consider the monodromy contribution. As shown in Figure
\ref{fig-cutoff}(b), we treat the first cusp in a special way. We
separate this cusp into two parts. One part is from $\delta_1$ at
the beginning, and the other part from the end of surface with
$\delta_{n+1}$, which includes the contribution of monodromy. We
choose half of each part so that to have an average contribution.
This is similar to the picture used in \cite{Alday:2009yn} for
$AdS_3$ case. The whole contribution is
\bea A_{\rm cutoff} = {1\over 2} \left[ \sum_{i=1}^{n} (L +
\delta_i)(L+ \delta_{i+1}) - {1\over2} \Delta_x \Delta_y +
(L+\delta_{n+1}) \Delta_y \right] . \eea
Notice that we need to take
\bea \delta_{n+1} = \delta_1 + \Delta_x ~, \qquad \delta_{n+2} =
\delta_2 + \Delta_y ~, \qquad \delta_{n+3} = \delta_3 - \Delta_x ~,
\qquad \textrm{and so on}. \eea

Now we need to solve for all $\delta_i$. Due to the monodromy
$\Delta_{x,y}$, it is no longer enough to consider only the
equations (\ref{eqell}). But there are many other equations as given
by (\ref{non-adjacent}), which involve non-adjacent kinematic
invariants
\bea \delta_i + \delta_j = \ell_{ij} - \log ({\hat X}_i\cdot {\hat
X}_j) ~. \eea
To solve our problem, it is enough to choose two of them, for
example
\bea && \delta_1 + \delta_4 = \ell_{14} - \log ({\hat X}_1\cdot {\hat X}_4)~, \label{ell14} \\
&& \delta_2 + \delta_5 = \ell_{25} - \log ({\hat X}_2\cdot {\hat
X}_5) ~. \label{ell25} \eea
The price is that we also introduce two new non-trivial variables
\bea {\hat X}_1\cdot {\hat X}_4 = \langle s_1 s_2 s_4 s_5 \rangle ~,
\qquad {\hat X}_2\cdot {\hat X}_5 = \langle s_2 s_3 s_5 s_6 \rangle
~.  \label{X1425}\eea
This two quantities are related to one of $T$ functions, $T_{2,1} =
\langle s_{-2} s_{-1} s_1 s_2 \rangle$. This $T$ functions can be
calculated from a limit of $Y$ function as we will show in next
section.

\vskip 0.1in

Therefore, the cutoff part is finally expressed in terms of
kinematic invariants $\ell_{ij}$ and $T$ functions. The terms that
related to the $T$ function will be defined as \textit{extra part}.
The remaining parts that only depend on kinematic invariants will be
defined as \textit{BDS-like part}. We will provide the explicit
expression of eight-point case in section \ref{section-8p}.

\section{$T$ function as a limit of $Y$ function \label{section-T}}

In this section, we calculate $T$ functions. We show that the $T$
functions can be obtained as a limit of $Y$ functions. The basic
idea is that we can obtain a lower-point structure by taking a limit
of a higher-point case. We will first show how to do this in the
$AdS_3$ case. The same calculation is then straightforward to
generalize to the $AdS_5$ case.
%
%

It is impossible to review the whole $Y$ system here. However, to
make the paper more self-contained, in particular to set up the
conventions, we provide a brief summary of $Y$ system in Appendix
\ref{appendix-TY}. Reader can find more details in
\cite{Alday:2010vh}

\subsection{The $AdS_3$ case}

\begin{figure}[t]
\centerline{\includegraphics[height=6.3cm]{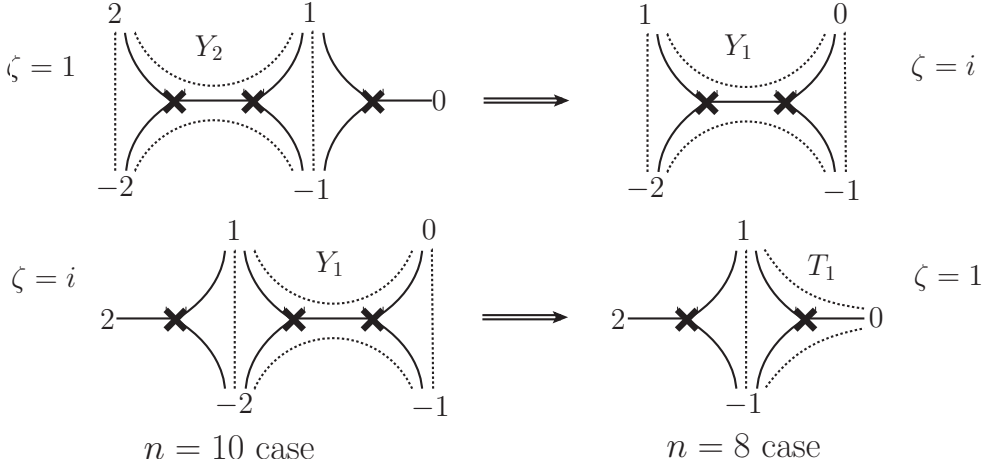} }
\caption{\it The limit behavior of the WKB pattern. The crosses are
zeros of the polynomial $p(z)$. The numbers indicate the various
Stokes sectors. The dotted lines are WKB lines which connect
different stokes sectors. The solid lines ending on the zeros
separate different classes of WKB lines. We consider two different
phases of $\zeta$, which show the contour formed by WKB lines for
$Y_2$ and $Y_1$ respectively. By taking the rightmost zero to
infinity, the structure of $n$=10 is reduced to that of $n$=8, and
$Y_2$ and $Y_1$ of the higher-point case are reduced to $Y_1$ and
$T_1$ of the lower-point case. Notice the change of labels of the
Stokes sectors in the limit.} \label{fig-ads3wkb}
\end{figure}

We focus on the function $T_{1}$, which will be related to the extra
part of the area. We start from two $Y$ functions (see
(\ref{Yads3}))
\bea \hat Y_{1} = Y_{1}^{-} = {\langle s_{-2} s_{1} \rangle \langle
s_{-1} s_0 \rangle\over \langle s_{-2} s_{-1} \rangle \langle s_0
s_1 \rangle} ~, \qquad \hat Y_{2} = Y_{2} = {\langle s_{-1} s_{1}
\rangle \langle s_{-2} s_{2} \rangle \over \langle s_{-2} s_{-1}
\rangle \langle s_1 s_2 \rangle} ~.\eea
The WKB lines corresponding to these two $Y$ function are shown in
Figure \ref{fig-ads3wkb}.
We consider the limit that the rightmost zero goes to infinity.
Notice the WKB lines (dotted lines) that connect different Stokes
sectors combine to form a contour which corresponds to a $Y$ or $T$
function as illustrated in the figure. By taking the rightmost zero
to infinity, we reduce the $n$-point structure to the structure of
($n$-2)-point. We can see explicitly that $Y_2$ and $Y_1$ of the
higher-point case are reduced to $Y_1$ and $T_1$ of a lower-point
case. Therefore we have that
\bea \big( \hat Y_{1}^{(n)} \big)^{+} & \rightarrow & {\langle
s_{-1} s_{1} \rangle \langle s_{0} s_0 \rangle\over \langle s_{-1}
s_0 \rangle \langle s_0 s_1 \rangle} \sim {\langle s_{-1} s_{1}
\rangle } = T_1^{(n-2)} ~,
\\ \hat Y_{2}^{(n)} & \rightarrow & \left( {\langle s_{-1} s_{0} \rangle
\langle s_{-2} s_{1} \rangle \over \langle s_{-2} s_{-1} \rangle
\langle s_0 s_1 \rangle} \right)^+ =  \big( \hat Y_1^{(n-2)} \big)^+
~, \label{limY2}\eea
The superscript of $Y^{(n)}$ means that it is a $Y$ function of the
$n$-point system. The ``+" can be obtained by considering the change
of the phase of $\zeta$.

One subtly here is that $\hat Y_1^{(n)}$ is actually vanishing in
this limit, due to the factor $\langle s_0 s_0 \rangle$. This is
because the limit of taking the zero to infinity actually
corresponds to the large $m_1$ limit, and since for large $m_s$ we
have
\bea \log Y_s = - m_s \cosh\theta + \cdots ~, \eea
$Y_1$ indeed goes to zero in the large $m_1$ limit. To evaluate the
$T_1$ function, we renormalize the $Y_1$ function by subtracting the
asymptotic WKB term $Y_{1,\textrm{\tiny WKB}} = e^{-m_s\cosh\theta}$
as
\bea T_1^{(n-2)} = \Big({ Y_1^{(n)} \over Y_{1,\textrm{\tiny
WKB}}^{(n)}}\Big)\Big|_{\rm m_1\rightarrow \infty}  ~. \eea
We mention that this is equivalent to making a choice for the
normalization of $T$ functions. Unlike $Y$ functions, $T$ functions
are not ``gauge invariant" due to the gauge redundancy of the Hirota
equation \cite{Alday:2010vh}. Therefore, such choice of
normalization is actually a choice of gauge fixing condition. A more
explicit discussion for this point and its relation to periods part
can be found in \cite{Yang:2010as}.

\vskip 0.1in

We can calculate $T_1$ now. As reviewed in Appendix
\ref{appendix-TY}, the general integral equations for $Y$ functions
are
\bea \log Y_s(\theta) = - m_s \cosh\theta + K \star \log(1+Y_{s+1})
(1+Y_{s-1}) ~, \qquad K(\theta) = {1\over 2\pi\cosh(\theta)} ~. \eea
For the first two $Y$ functions we have
\bea \log Y_1 &=& - m_1 \cosh\theta + K \star \log(1+Y_2) ~, \\
\log Y_2 &=& - m_2 \cosh\theta + K \star \log(1+Y_1) + K \star
\log(1+Y_3) ~. \eea
In the large $m_1$ limit, we have
\bea \log T_1^{(n-2)} = \log \Big({Y_1^{(n)} \over
Y_{1,\textrm{\tiny WKB}}^{(n)}}\Big)\Big|_{\rm m_1\rightarrow
\infty} = K \star \log(1+Y_2^{(n){\rm lim}}) ~. \label{y2lim} \eea
We also have
\bea \log Y_2^{(n),{\rm lim}} = - m_2 \cosh\theta + K \star
\log(1+Y_3^{(n),{\rm lim}}) ~, \eea
which is in a integral form of $Y_1$, which is consistent with the
(\ref{limY2}) that $Y_2^{(n)} \rightarrow Y_1^{(n-2)}$ (and more
generally $Y_s^{(n)} \rightarrow Y_{s-1}^{(n-2)}$). So we have the
$T_1$ function in the ($n$-2)-point system as
\bea \log T_1 = K \star \log(1+ Y_1) ~. \eea
This may be written in a more explicit form as (for $\varphi_1 \in
(-\pi/2, \pi/2)$)
\bea \log T_1 = {1\over2\pi} \int_{-\infty}^\infty d \theta' {1\over
\cosh(\theta'-\theta+i\varphi_1)} \log \left( 1 + Y_1 (\theta')
\right) ~. \eea
For the case of $n=8$, this is exactly the same expression of as
(6.5) in \cite{Alday:2009yn}:
\bea && \log \gamma_1 = {1\over2\pi} \int_{-\infty}^\infty d \theta'
{1\over \cosh(\theta'-\theta+i\varphi)} \log \left( 1 + e^{-|m|
\cosh\theta'} \right) ~, \eea
where $\gamma_1$ was called ``Stokes parameter" there%
\footnote{Notice one should replace $m$ with $m/2\pi$ for the result
in \cite{Alday:2009yn} to accord with the convention here.}. This
shows that the Stokes parameter can actually be understood as $T$
function. Our definition of the extra part is therefore the same as
the definition in \cite{Alday:2009yn} which is related to the Stokes
parameter.

\subsection{The general $AdS_5$ case}

The above prescription can be directly generalized to the $AdS_5$
case. To calculate the extra part we need to calculate $T_{2,1}$ as
shown below (\ref{X1425}). We consider $Y_{a,s}$ for $s=1,2$ which
may be written explicitly as
\bea \hat Y_{2,1} & \equiv &  Y_{2,1}^- = {\langle -3,-2,1,2 \rangle
\langle -2,-1,0,1 \rangle \over \langle -3,-2,-1,1 \rangle \langle
-2,0,1,2 \rangle} ~, \\ \hat Y_{2,2} & \equiv &  Y_{2,2} = {\langle
-3,-2,2,3 \rangle \langle -2,-1,1,2 \rangle \over \langle -3,-2,-1,2
\rangle \langle -2,1,2,3 \rangle} ~, \\
\hat Y_{1,1} & \equiv & Y_{1,1} = {\langle -1,0,1,2 \rangle \langle
-2,1,2,3 \rangle
\over \langle 0,1,2,3 \rangle \langle -2,-1,1,2 \rangle} ~, \\
\hat Y_{1,2} & \equiv &  Y_{1,2}^- = {\langle -4,-3,-2,2 \rangle
\langle -3,-2,-1,1 \rangle \over \langle -4,-3,-2,-1 \rangle \langle
-3,-2,1,2 \rangle} ~, \\
\hat Y_{3,1} & \equiv &  Y_{3,1} = {\langle -2,-1,0,1 \rangle
\langle -3,-2,-1,2 \rangle \over \langle -2,-1,1,2 \rangle \langle
-3,-2,-1,0 \rangle} ~, \\ \hat Y_{3,2} & \equiv &  Y_{3,2}^- =
{\langle -3,1,2,3 \rangle \langle -2,0,1,2 \rangle \over \langle
-3,-2,1,2 \rangle \langle 0,1,2,3 \rangle} ~. \eea
In exactly the same picture as the $AdS_3$ case, we take the zero in
the rightmost side to infinity and obtain that
\bea \big(\hat Y_{2,1}^{(n)} \big)^+ &\rightarrow& {\langle
-2,-1,1,2 \rangle \langle -1,0,0,1 \rangle \over \langle -2,-1,0,1
\rangle \langle
-1,0,1,2 \rangle} \sim \langle -2,-1,1,2 \rangle = T_{2,1}^{(n-1)} ~, \label{limitY21} \\
\hat Y_{2,2}^{(n)} &\rightarrow& \left ({\langle -3,-2,1,2 \rangle
\langle -2,-1,0,1 \rangle \over \langle -3,-2,-1,1 \rangle \langle
-2,0,1,2 \rangle} \right)^+ = \big( \hat
Y_{2,1}^{(n-1)} \big)^+ = Y_{2,1}^{(n-1)} ~, \\
\hat Y_{1,1}^{(n)} &\rightarrow& \left( {\langle -1,0,0,1 \rangle
\langle -2,0,1,2 \rangle \over \langle 0,0,1,2 \rangle \langle
-2,-1,0,1 \rangle}\right)^+ \sim \left(\langle
-2,0,1,2 \rangle \right)^+ = T_{1,1}^{(n-1)} ~, \\
\big( \hat Y_{1,2}^{(n)} \big)^+ &\rightarrow& {\langle -2,-1,0,1
\rangle \langle -3,-2,-1,2 \rangle \over \langle -2,-1,1,2 \rangle
\langle
-3,-2,-1,0 \rangle} = Y_{3,1}^{(n-1)} ~, \\
\hat Y_{3,1}^{(n)} &\rightarrow& \left( {\langle -2,-1,0,0 \rangle
\langle -3,-2,-1,1 \rangle \over \langle -2,-1,0,1 \rangle \langle
-3,-2,-1,0 \rangle} \right)^+ \sim \left(\langle -3,-2,-1,1 \rangle \right)^+ = T_{3,1}^{(n-1)} ~, \\
\big( \hat Y_{3,2}^{(n)} \big)^+ &\rightarrow& {\langle -1,0,1,2
\rangle \langle -2,1,2,3 \rangle \over \langle 0,1,2,3 \rangle
\langle -2,-1,1,2 \rangle} = Y_{1,1}^{(n-1)} ~. \label{limitY32}
\eea
There are also some terms go to zero in the limit for $\hat
Y_{a,1}$. Similar to $AdS_3$ case, we can renormalize them by
subtracting the WKB term as
\bea T_{2,1}^{(n-1)} = \left({Y_{2,1}^{(n)} \over
Y_{{2,1}\textrm{\tiny WKB}}^{(n)}}\right)\Big|_{\rm m_1\rightarrow
\infty}  ~, \eea
where $Y_{{2,1},\textrm{\tiny WKB}} = e^{-\sqrt{2}m_1\cosh\theta}$.
Using the integral form of $Y$ functions (\ref{logY2s}), we obtain
\bea \log T_{2,1}^{(n-1)} &=& \log \left({ Y_{2,1}^{(n)} \over
Y_{{2,1}\textrm{\tiny WKB}}^{(n)}}\right)\Big|_{\rm m_1\rightarrow
\infty} \nonumber\\ &=& K_2 \star \log (1+Y_{2,2}^{(n)\rm lim}) +K_1
\star \log (1+Y_{1,2}^{(n)\rm lim})(1+Y_{3,2}^{(n)\rm lim}) ~, \eea
From (\ref{limitY21})-(\ref{limitY32}), we know that $Y_{2,2}^{(n)},
Y_{1,2}^{(n)}, Y_{3,2}^{(n)}$ is reduced to $Y_{2,1}^{(n-1)},
Y_{3,1}^{(n-1)}, Y_{1,1}^{(n-1)}$ in the limit, therefore we obtain
the $T_{2,1}$ function in a ($n$-1)-point system as
\bea \log T_{2,1} & = &  K_2 \star \log (1+ Y_{2,1}) +K_1 \star \log
(1+Y_{1,1})(1+Y_{3,1}) ~. \label{T21}\eea
We can also derive the formula for $T_{1,1}$ and $T_{3,1}$ in the
same way. The finally expressions are
\bea \log T_{1,1} & = &  {1\over2} K_2 \star \log (1+ Y_{1,1}) (1+
Y_{3,1}) + K_1 \star \log (1+Y_{2,1}) - {1\over2} K_3 \star
\log {(1+ Y_{1,1}) \over (1+ Y_{3,1})} ~, \\
\log T_{3,1} & = & {1\over2} K_2 \star \log (1+ Y_{1,1}) (1+
Y_{3,1}) + K_1 \star \log (1+Y_{2,1}) + {1\over2} K_3 \star \log
{(1+ Y_{1,1}) \over (1+ Y_{3,1})} ~. \eea

It is easy to check these results indeed yield the required
functional relations (similar to what was done for $Y$ function in
\cite{Alday:2010vh})
\bea T_{1,1}^+ T_{3,1}^- = T_{2,1}(1+Y_{3,1}) ~, \quad  T_{3,1}^+
T_{1,1}^- = T_{2,1}(1+Y_{1,1}) ~, \quad T_{2,1}^+ T_{2,1}^- =
T_{1,1}T_{3,1}(1+Y_{2,1}) ~, \eea
by using the identities of kernels that
\bea K_2^+ + K_2^- = \delta(\theta) + 2K_1 ~, \quad K_1^+ + K_1^- =
K_2 ~, \quad K_3^+ - K_3^- = -\delta(\theta) ~. \eea
In the above integral form, the $T$ functions can be calculated in
the same way as $Y$ functions.

\section{A conjecture for periods part \label{section-periods}}

The final missing piece is the periods part. As we mentioned in
section \ref{section-amplitudes}, the periods part is the difference
between the surface $\Sigma$ and simplified surface $\Sigma_0$. It
contains the branch cut information which depends on the polynomial
$P(z)$. It also depends on how we choose the surface $\Sigma_0$. For
the $n\neq 4K$ case, the $\Sigma_0$ surface can be defined by
choosing a simple polynomial $P_0(z) = z^{n-4}$. Periods part can
then be given explicitly as
\bea A_{\rm periods}^{n\neq 4K} = 2 \int d^2 z \left( [ P(z) \bar
P(\bar z)] ^{1/4} - |z|^{n/2-2} \right) ~. \eea
For the case that $n=4K$, due to the monodromy, we cannot choose
such a simple polynomial for $\Sigma_0$.

As we mentioned before, the periods part is expressed in terms of
periods around cycles of the Riemann surface. The corresponding
Riemann surface for $AdS_5$ case is defined as $x^4$=$P(z)$ which is
a quadruple cover of Riemann sphere. While for $AdS_3$ case, it is a
simpler double branch cover given as $x^2$=$p(z)$, and the periods
part was given in \cite{Alday:2009yn}. We will first review the
result of $AdS_3$ case, and then make a direct generalization to
$AdS_5$ case.

\vskip 0.1in

\begin{figure}[t]
\centerline{\includegraphics[height=4cm]{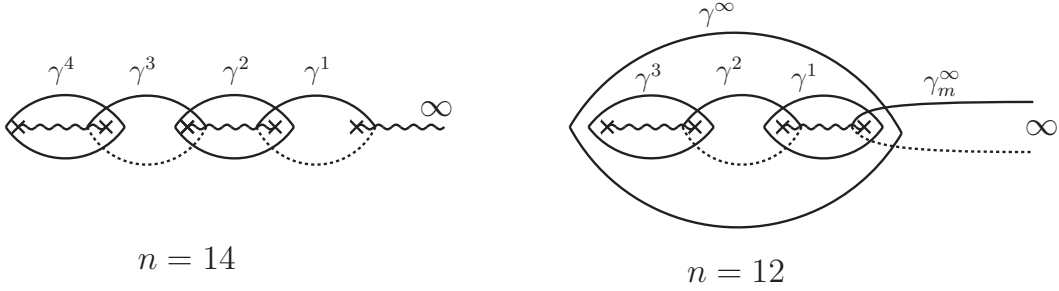} }
\caption{\it The pattern of cycle structure for the Riemann surface.
The crosses represent the zeros of polynomial $p(z)$. The wave lines
indicate the branch cuts. There is a branch point at infinity for
$n=4K+2$ case. Notice the the cycle $\gamma_m^\infty$ in $n=12$ may
be taken as the the cycle $\gamma^1$ in $n=14$ by taking the
rightmost zero to infinity.} \label{fig-cycle}
\end{figure}

To study the periods part, we need to choose a basis of cycle for
the Riemann surface. Following \cite{Alday:2009yn}, we choose
\bea n = 4K+2 : && \gamma^s ~, \quad s=1, \cdots , {n-6\over2} ~, \\
n = 4K : && \gamma^s ~, \quad s=2, \cdots , {n-8\over2} ~, \qquad
{\rm and}~ \gamma^\infty ~, ~\gamma_m^{\infty} ~, \eea
where the case of $n=14$ and $n=12$ are shown explicitly in Figure
\ref{fig-cycle}. Other cases have similar patterns.

We can see that while all cycles are compact for $n=4K+2$ case,
there is a non-compact cycle $\gamma_m^\infty$ when $n=4K$. Its dual
cycle $\gamma^\infty$ goes around infinity, over which the
integration gives the monodromy we discussed before. As mentioned in
\cite{Alday:2009yn}, to obtain the correct normalization of Stokes
parameter (i.e. the $T$ function), we need to choose the origin of
$w$-plane to be the zero which the non-compact cycle is around. The
results of periods part are given in \cite{Alday:2009yn}, which
involves only non-infinite cycles. The final expression of the
periods part turns out to have the same expression as $n=4K-2$ case%
\footnote{One may understand this point by considering that the
information of the two infinite cycles, which is related to the
monodromy, is already included in the cutoff part.}. The periods
part can therefore be explicit defined as
\bea A_{\rm periods}^{n=4K} = 2 \int d^2 z \left( [ \tilde p(z) \bar
{\tilde p}(\bar z)] ^{1/2} - |z|^{{n-2\over2}-2} \right) ~, \qquad ~
\tilde p(z) = {p(z)\over z-z_1} ~, \eea
where $z_1$ is the zero in the most right side, which both
$\gamma^\infty$ and $\gamma_m^\infty$ go around, and is defined as
the origin of the $\Sigma_0$ surface. For the $n=8$ case in $AdS_3$,
the periods part is same as six-point case and therefore should be
zero. This is indeed true, for example let $p(z)= z^2 - a^2$ and
$\tilde p(z) = z + a$, we have
\bea A_{\rm periods}^{n=8,{AdS_3}} = 2 \int d^2 z \left
(\sqrt{(z+a)(\bar z+\bar a)} - \sqrt{z\bar z} \right) = 0 ~. \eea

Following the above picture, we may conjecture the periods part in
$AdS_5$ case to be similarly given as
\bea A_{\rm periods}^{n=4K, {\rm conjecture}} = 2 \int d^2 z \left(
[ \tilde P(z) \bar {\tilde P}(\bar z)] ^{1/4} - |z|^{{n-2\over2}-2}
\right) ~, \qquad ~ \tilde P(z) = {P(z)\over z-z_1} ~, \eea
where under this assumption, the $4K$-point result should have the
same expression as the $(4K-1)$-point case by replacing $m_s$ with
$m_{s+1}$, for example for eight-point we may obtain from the result
of seven points as
\bea A_{\rm periods}^{n=8, {\rm conjecture}} = {|m_2|^2 + |m_3|^2
\over 2} + {m_2 \bar m_3 + \bar m_2 m_3 \over 2 \sqrt{2}} ~. \eea
Since the structure of Riemann surfaces is more complicated than
$AdS_3$ case, it may not be surprising if there are extra
contribution in $AdS_5$ case. It would be important to check whether
this
generalization is correct or not%
\footnote{Note added: The periods parts are calculated in a later
paper by the author by considering a special collinear limit
\cite{Yang:2010as}. In this limit, the periods part can be uniquely
fixed by the BDS part (the one-loop finite part of amplitudes at
weak coupling\cite{BDS}), and the cutoff part calculated in this
paper. There is indeed extra contribution compared to the conjecture
here. We cite the correct periods part of eight-point in next
section (\ref{8periods}).}.

\section{Eight-point result \label{section-8p}}

In this section we present the explicit calculation for eight-point
case. The cutoff part is given as
\bea && A_{\rm cutoff} = A_{\rm div} + A_{\rm fincut} ~, \\
&& A_{\rm div} = {1\over 2} \sum_{i=1}^8 \Big( L+{\delta_i +
\delta_{i+2} \over2} \Big)^2 ~,
\\ && A_{\rm fincut}
= -{1\over 8} \sum_{i=1}^8 \ell_i^2 +
{1\over 2} \sum_{i=1}^{n} \delta_i \delta_{i+1} + {1\over 4}
\Delta_x \Delta_y + {1\over 2} \delta_1 \Delta_y ~. \eea
where $A_{\rm fincut}$ contains both BDS-like and extra part. Notice
$\delta_9 = \delta_1 + \Delta_x, \delta_{10} = \delta_2 + \Delta_y$.
We can solve $\delta_i$ plus $\Delta_{x,y}$ from equations
(\ref{eqell}), (\ref{ell14}) and (\ref{ell25}), which we collected
here
\bea && \delta_i + \delta_{i+2} = \ell_i \equiv \log x_{i,i+2}^2
~,\\
&& \delta_1 + \delta_4 = \ell_{14} - \log T_{2,1}^{[6]} ~,
\\ && \delta_2 + \delta_5 = \ell_{25} - \log T_{2,1}^{[8]} ~, \eea
where we have used the relation $T_{2,1}^{[6]}=\langle s_1 s_2 s_4
s_5 \rangle$ and $T_{2,1}^{[8]}=\langle s_2 s_3 s_5 s_6 \rangle$. We
write the monodromy explicitly
\bea \Delta_x = -\ell_1+\ell_3-\ell_5+\ell_7 = \log \langle s_1 s_2
s_7 s_8 \rangle ~, \qquad \Delta_y = -\ell_2+\ell_4-\ell_6+\ell_8 =
\log \langle s_2 s_3 s_8 s_9 \rangle ~. \eea
where we have used (\ref{deltaT}).

The parts related to $T$ functions are define as extra part
\bea A_{\rm extra} = -{1\over4} \left[ (\Delta_x+\Delta_y) \log
T_{2,1}^{[6]} - (\Delta_x-\Delta_y) \log T_{2,1}^{[8]} \right] ~,
\label{extra-1} \eea
where $T_{2,1}$ function can be calculated using (\ref{T21}). Notice
that to calculate $T_{2,1}^{[6,8]}$, one needs to generalize the
equation (\ref{T21}) to other phase regions of $\zeta$-plane, where
pole terms should be included \cite{Alday:2010vh}. The BDS-like part
is given by the remaining parts as
\bea A_{\rm BDS-like} &=& A_{\rm fincut} - A_{\rm extra} \nonumber\\
&=& {1\over 8} \sum_{i=1}^{8} \big[ \ell_i^2 - (\ell_i
-\ell_{i+1})^2 \big] + {1\over 4} \big[\ell_1 \Delta_x + \ell_2
\Delta_y - \ell_3 \Delta_x
- \ell_4 (\ell_3+\ell_7) - \ell_8(\ell_1+\ell_5) \big] \quad \nonumber\\
&& + {1\over4} \big[ (\Delta_x+\Delta_y) \ell_{14} -
(\Delta_x-\Delta_y) \ell_{25} \big] ~. \label{bdslike-1} \eea
Other parts of the amplitude are%
\footnote{We have cited the results of periods part obtained in
\cite{Yang:2010as}, with which the amplitude has the correct
collinear limit.}
\bea A_{\rm periods} = {|m_2|^2 + |m_3|^2 \over 2} + {m_2 \bar m_3 +
\bar m_2 m_3 \over 2 \sqrt{2}} - {1\over 4}|m_1 + \sqrt{2}m_2 +
m_3|^2 ~, \label{8periods} \eea
\bea A_{\rm free} = \sum_{s=1}^3 { |m_s| \over 2\pi}
\int_{-\infty}^{+\infty} {d\theta }  \cosh\theta \log \left[(1+
Y_{1,s})(1+ Y_{3,s})(1+ Y_{2,s})^{\sqrt{2}}\right] ~. \eea
The whole area is (up to a constant)
\bea A &=& A_{\rm div} + A_{\rm BDS-like} + A_{\rm extra} + A_{\rm
periods} + A_{\rm free} ~. \eea

\subsection{Another choice of equations}

To calculate $\delta_i$, we have chosen two extra conditions
(\ref{ell14}) and (\ref{ell25}). We may choose other equations as
well. For example
\bea && \delta_8 + \delta_4 ~=~ \log x_{48}^2 - \log ({\hat
X^+}_8\cdot {\hat X^+}_4) ~=~ \ell_{48} - \log T_{2,2}^{[5]}~,
\\ && \delta_7 + \delta_3 ~=~
\log x_{37}^2 - \log ({\hat X^+}_7\cdot {\hat X^+}_3) ~=~ \ell_{37}
- \log T_{2,2}^{[3]} ~. \eea
We obtain a different expression for BDS-like and extra part
\bea A_{\rm BDS-like} &=& -{1\over 8} \sum_{i=1}^{8} \ell_i^2  +
{1\over 4} \sum_{i=1}^{8} \ell_i \ell_{i+1} - {1\over 4} (\ell_2 +
\ell_6) (\ell_3 + \ell_7) + {1\over 4}\left( \Delta_x \ell_{48}  -
\Delta_y \ell_{37} \right)  ~, \label{bdslike-2} \\
A_{\rm extra} &=& -{1\over 4}\left( \Delta_x \log T_{2,2}^{[-3]} -
\Delta_y \log T_{2,2}^{[-5]} \right) ~, \label{extra-2} \eea
With such choices we have to evaluate the $T_{2,2}$ function
$\langle s_i s_{i+1} s_{i+4} s_{i+5}\rangle$, which can be
calculated by using the relation $T_{2,2} = T_{1,1} T_{3,1}
Y_{2,1}$.

We can see that expression of BDS-like and extra parts depends on
the choice of equations. There is no unique definition for each of
them. However, the summation of extra and BDS-like part must be
invariant. We can check this explicit.

The difference between the BDS-like part is
\bea (\ref{bdslike-1}) - (\ref{bdslike-2}) &= & {1\over4}
(\ell_1+\ell_5 -
\ell_3-\ell_7)(\ell_1+\ell_{2,5}+\ell_{4,8}-\ell_3-\ell_8 -
\ell_{1,4}) \nonumber\\ && + {1\over4} (\ell_2+\ell_6 -
\ell_4-\ell_8)(\ell_2+\ell_3+\ell_7+\ell_{1,4}+\ell_{2,5} -
\ell_{3,7}) \nonumber\\ &=& - {1\over4} \Delta_x \log \left(
{x_{13}^2 x_{25}^2 x_{48}^2 \over x_{35}^2 x_{82}^2 x_{14}^2}
\right) - {1\over4} \Delta_y \log \left( {x_{24}^2 x_{35}^2 x_{71}^2
\over x_{14}^2 x_{25}^2 x_{37}^2} \right) \nonumber\\ &=& -
{1\over4} \Delta_x \log \left( {\langle s_2 s_3 s_5 s_6 \rangle
\langle s_4 s_5 s_8 s_9 \rangle \over \langle s_1 s_2 s_4 s_5
\rangle} \right) + {1\over4} \Delta_y \log \left( {\langle s_1 s_2
s_4 s_5 \rangle \langle s_2 s_3 s_5 s_6 \rangle \langle s_3 s_4 s_7
s_8 \rangle} \right) ~, \nonumber \eea
This exactly cancels the difference between the extra parts
$(\ref{extra-1})-(\ref{extra-2})$, by noticing that
\bea T_{2,1}^{[6]} = \langle s_1 s_2 s_4 s_5 \rangle ~, \quad
T_{2,1}^{[8]}  = \langle s_2 s_3 s_5 s_6 \rangle  ~, \quad
T_{2,2}^{[-3]} = \langle s_4 s_5 s_8 s_9 \rangle ~, \quad
T_{2,2}^{[-5]} = \langle s_3 s_4 s_7 s_8 \rangle ~. \eea
Therefore, the cutoff parts with two different choices of equations
are indeed equivalent to each other.

\section{Discussion \label{section-dis}}

Let us make more comments on the BDS-like part. First notice that
the BDS-like part (plus the universal divergent part) gives the
correct dual conformal anomaly, since all other parts are conformal
invariant. Therefore the difference between BDS-like and BDS part
must be a conformal invariant functions. As we have seen that in
$n$=$4K$ case, we do not have a unique definition of BDS-like part.
This is different from the $n\neq 4K$ case, in which the BDS-like
part is uniquely expressed in terms of only adjacent kinematic
invariants. The reason for this uniqueness is that, in $n\neq 4K$
case, we cannot express any cross ratios in terms of only adjacent
kinematic invariants, therefore the expression is fixed by conformal
Ward identity.

We may explain this point more explicitly. Suppose we can express
the BDS-like part in terms of only adjacent kinematic invariants
$\ell_i$, i.e. $A_{\rm BDS-like} = F(\ell_i)$, which gives correct
conformal anomaly. The expression will not be unique if we can also
express some function of cross ratios $u$ in terms of only $\ell_i$,
for example $g(u)=f(\ell_i)$. This is because we can define a new
function $A'_{\rm BDS-like} = F(\ell_i) + f(\ell_i)$ which also
satisfies the Ward identity. In the $n\neq 4K$ case, we cannot have
any relation as $g(u)=f(\ell_i)$, therefore the function $F(\ell_i)$
is uniquely fixed by Ward identity. But when $n=4K$, we do have such
relations as $g(u)=f(\ell_i)$. At the same time, it is also
impossible to have a function $F(\ell_i)$ which can give correct
anomaly. Therefore, non-adjacent kinematic invariants are necessary
to appear in the final expression. We have seen this explicitly,
since we must introduce two new equations for calculating
$\delta_i$, which involve non-adjacent kinematic invariants.

Besides the choice of equations,  we also made several other choices
during the calculation. When computing the cutoff area, we treated
the first cusp in a special way. We make some gauge choice which is
related to the normalization of $T$ functions. While considering the
periods part, we choose the origin of $w$-plane to be one of the
zeros of the polynomial, which is also implicitly related to the
normalization of $T$ functions. Of course, the physics i.e. the
whole result should be independent of all these choices, as we have
checked for the cutoff part. We emphasize that we only make a
conjecture for the periods part in this paper, and it would be
important to calculate this part more honestly and check the
conjecture.

Finally, we mention that there are other important open problems, of
which the most challenging one is perhaps how to calculate the
amplitude at arbitrary value of 't Hooft coupling constant. One can
expect the quantum integrability \cite{Kazakov:2004qf,
Arutyunov:2004vx} should play an essential role to realize this. It
would also be interesting to study and see if we can apply these
method to study the $S$-matrix in a cousin of ${\cal N}=4$ SYM, the
ABJM theory \cite{ABJM}. Some observations for amplitude and Wilson
loop duality at weak coupling side are given in \cite{Henn:2010ps,
Bargheer:2010hn}.

\section*{Acknowledgements}
The author would like to thank Andreas Brandhuber and Gabriele
Travaglini for discussion and collaboration on a related topic which
inspired the study of the present problem. He is very grateful to
Gabriele Travaglini for his carefully reading the draft and
suggestions for the presentation. He would especially like to thank
Andreas Brandhuber for the encouragement and drawing the author's
attention back to the very interesting strong coupling story. It is
also a pleasure to thank Wei Song for her helpful comments on the
draft. This work is supported by the STFC.

\appendix

\section{A brief summary of $Y$ system \label{appendix-TY}}

Here we summarize the main result of $Y$ system, and at the same
time set up the convention. Reader can find details in
\cite{Alday:2010vh}.

\subsection{$AdS_3$ case}

We use the convention that (the convention will be different in
$AdS_5$ case)
\bea f^\pm(\zeta) = f(e^{\pm i\pi/2} \zeta) ~, \qquad f^{[m]}(\zeta)
= f(e^{i m\pi/2} \zeta) ~. \eea

$T$ functions are defined as
\bea && T_{1,2k+1} = \langle s_{-k-1} s_{k+1}\rangle~, \qquad
T_{1,2k} = \langle s_{-k-1} s_k \rangle^+ ~, \\  && T_{0,2k} =
\langle s_{-k-1} s_{-k}\rangle ~, \qquad
T_{0,2k+1} = \langle s_{-k-2} s_{-k-1} \rangle^+ ~, \\
&& T_{2,2k} = \langle s_{k} s_{k+1} \rangle ~, \qquad\quad
T_{2,2k+1} = \langle s_{k} s_{k+1} \rangle^+ ~, \eea
where the contraction of smallest solution is defined as $\langle
s_i s_j \rangle = \epsilon_{\alpha\beta} s_i^\alpha s_j^\beta$.

$Y$ functions are defined as
\bea Y_{m} = {T_{1,m-1} T_{1,m+1} \over T_{0,m} T_{2,m}} ~.
\label{Yads3} \eea

We can choose the normalization conditions
\bea \langle s_i s_{i+1} \rangle = 1 ~, \eea
which is the gauge fixing conditions for $T$ functions. We also have
the shifting relation
\bea \langle s_i s_j \rangle^{[2]} = \langle s_{i+1} s_{j+1} \rangle
~, \eea
which comes from the $Z_2$ symmetry of the corresponding $SU(2)$
Hitchin system.

The $Y$ functions satisfy the functional relations
\bea Y_s^- Y_s^+ = (1+Y_{s-1})(1+Y_{s+1}) ~, \qquad s = 1,2,\cdots,
{n\over2}-3 ~. \eea
The equivalent integral form can be given as
\bea \log Y_s(\theta) = - m_s \cosh\theta + K \star \log(1+Y_{s+1})
(1+Y_{s-1}) ~, \qquad K(\theta) = {1\over 2\pi\cosh\theta} ~.
\label{logYs} \eea
Notice that in this form we assume the phase $\varphi_s$ of $m_s$ to
be zero, and the valid range for the phase of $\zeta$ (or the
imaginary part of $\theta$) is $\phi \in (-\pi/2, \pi/2)$.

\subsection{$AdS_5$ case}

We use a different convention that
\bea f^\pm(\zeta) = f(e^{\pm i\pi/4} \zeta) ~, \qquad f^{[m]}(\zeta)
= f(e^{i m\pi/4} \zeta) ~. \eea

$T$ functions are defined as
\bea T_{0,m} = \langle s_m s_{m+1} s_{m+2} s_{m+3} \rangle^{[-m-1]}
~, \qquad T_{4,m} = \langle s_{-2} s_{-1} s_0 s_1 \rangle^{[-m-1]}
~, \eea
\bea T_{1,m} = \langle s_{-2} s_{-1} s_0 s_{m+1} \rangle^{[-m]} ,
\quad T_{2,m} = \langle s_{-1} s_{0} s_{m+1} s_{m+2}
\rangle^{[-m-1]} , \quad T_{3,m} = \langle s_{-1} s_{m} s_{m+1}
s_{m+2} \rangle^{[-m]} . \label{Tads5} \nonumber\\\eea
where the contraction is defined as $\langle s_i s_{i+1} s_j s_{j+1}
\rangle \equiv \epsilon^{\alpha\beta\gamma\delta} s_{i,\alpha}
s_{i+1,\beta} s_{j,\gamma} s_{j+1,\delta}$.

$Y$ functions are defined as
\bea Y_{a,m} = {T_{a,m-1} T_{a,m+1} \over T_{a-1,m} T_{a+1,m}} ~.
\eea

We have the normalization condition that
\bea \langle s_i s_{i+1} s_{i+2} s_{i+3} \rangle = 1 ~,
\label{s1234} \eea
and the $Z_4$ symmetry provides the shifting relations
\bea && \langle s_{j-1} s_j s_{k-1} s_k \rangle^{[2]} = \langle s_j
s_{j+1} s_k s_{k+1} \rangle ~, \\ && \langle s_{j-2} s_{j-1} s_j s_k
\rangle^{[2]} = \langle s_j s_k s_{k+1} s_{k+2} \rangle ~, \\ &&
\langle s_j s_{k-2} s_{k-1} s_k \rangle^{[2]} = \langle s_j s_{j+1}
s_{j+2} s_k \rangle ~. \eea

The $Y$ functions satisfy the following functional relations
\bea { Y_{2,s}^- Y_{2,s}^+ \over Y_{1,s} Y_{3,s} } & = & {
(1+Y_{2,s+1}) (1+Y_{2,s-1}) \over (1+Y_{1,s})(1+Y_{3,s})} ~, \\
{ Y_{3,s}^- Y_{1,s}^+ \over Y_{2,s} } & = & {
(1+Y_{3,s+1}) (1+Y_{1,s-1}) \over 1+Y_{2,s}} ~, \\
{ Y_{1,s}^- Y_{3,s}^+ \over Y_{2,s} } & = & { (1+Y_{1,s+1})
(1+Y_{3,s-1}) \over 1+Y_{2,s}} ~, \eea
where for $n$-point, $s=1,2,...,n-5$. The equivalent integral form
is
\bea \log Y_{2,s} & = & - \sqrt{2} m_s \cosh\theta - K_2 \star
\alpha_s - K_1 \star \beta_s ~,  \label{logY2s} \\ \log Y_{1,s} & =
& - m_s \cosh\theta - C_s - {1\over2} K_2 \star \beta_s - K_1 \star
\alpha_s
- {1\over2} K_3 \star \gamma_s ~, \\
\log Y_{3,s} & = & - m_s \cosh\theta + C_s - {1\over2} K_2 \star
\beta_s - K_1 \star \alpha_s + {1\over2} K_3 \star \gamma_s ~, \eea
where
\bea && \alpha_s \equiv \log {(1+Y_{1,s})(1+Y_{3,s}) \over
(1+Y_{2,s-1})(1+Y_{2,s+1}) } ~, \qquad \gamma_s \equiv \log
{(1+Y_{1,s-1})(1+Y_{3,s+1}) \over (1+Y_{1,s+1})(1+Y_{3,s-1}) } ~, \\
&& \beta_s \equiv \log { (1+Y_{2,s})^2 \over
(1+Y_{1,s-1})(1+Y_{1,s+1})(1+Y_{3,s-1})(1+Y_{3,s+1})} ~, \eea
and
\bea K_1 \equiv {1 \over 2 \pi} {1\over\cosh \theta} ~, \qquad  K_2
\equiv {\sqrt{2} \over \pi} {\cosh\theta\over\cosh 2\theta} ~,
\qquad K_3 \equiv {i \over \pi} {\tanh 2\theta} ~. \eea
%


\end{document}